\documentclass[]{aa}
\usepackage{graphicx}
\usepackage{amssymb,amsmath}
\usepackage{txfonts}
\usepackage{longtable,lscape}
 \usepackage{epsfig}
 \usepackage{natbib}
\usepackage{epsfig}
\usepackage[normalem]{ulem}

\begin{document}

\def\kms {$\rm{km~s^{-1}}$}
\def\kmsmpc {\rm{km~s^{-1}}~Mpc^{-1}~}
\def\masa{h^{-1}{{\cal M}_{\odot}}~}
\def\hbeta{H$\beta$}
\def\hdelta{H$\delta$}
\def\halfa{H$\alpha$}
\def\hgama{H$\gamma$}

\def\lsigma{$L-\sigma$~}
\def\lsigmaw{$L-\sigma-EW(H\beta)$~}
\def\eqw{\rm EW(H$\beta$)}
\def\lbeta{L(H$\beta$)~}
\def\lhbeta{L(H$\beta$)~}
\def\xigma{$\sigma$~}
\def\vdisp{$\sigma$~}
 
\def\apjs   {Astro\-phys.\nobreak\ J.\ Suppl.\nobreak\ }
\def\aaps {Astr.\ Astro\-phys.\nobreak\ Suppl.\nobreak\ }
\def\mnras  {Mon.\ Not.\ R.\ astr.\nobreak\ Soc.\nobreak\ }
\def\apj   {Astro\-phys.\nobreak\ J.\nobreak\ }
\def\aanda  {Astr.\ Astro\-phys.\nobreak\ }
\def\aap  {Astr.\ Astro\-phys.\nobreak\ }
\def\apss {Astro\-phys.\nobreak\ Sp.\nobreak\ Sc.\nobreak\ }
\def\aj     {Astron.\nobreak\ J.\nobreak\ }

\title{The \lsigma relation for HII galaxies in green }

\author{ J.~Melnick\inst{1,2}, E.~Telles\inst{2}, V.~Bordalo\inst{2}, R. Ch\'avez\inst{3,4}, D.~Fern\'andez-Arenas\inst{5},
E.~Terlevich\inst{5}, R.~Terlevich\inst{5,6}, F.~Bresolin\inst{7}, M.~Plionis\inst{8}, S.~Basilakos\inst{9} }
\institute{European Southern Observatory, Av. Alonso de Cordova 3107, Santiago, Chile,
\and
Observatorio Nacional, Rua Jos\'e Cristino 77, 20921-400 Rio de Janeiro, Brasil,
\and
Cavendish Laboratory, University of Cambridge, 19 J. J. Thomson Ave, Cambridge CB3 0HE, UK, 
\and
Kavli Institute for Cosmology, University of Cambridge, Madingley Road, Cambridge CB3 0HA, UK, 
\and
Instituto Nacional de Astrof{\'i}sica {\'O}ptica y Electr{\'o}nica, AP 51 y 216, 72000, Puebla, M{\'e}xico,
\and
Institute of Astronomy, University of Cambridge, Madingley Road, Cambridge CB3 0HA, UK,
\and
Institute for Astronomy of the University of Hawaii, 2680 Woodlawn Drive, 96822 Honolulu, HI, USA
\and
Physics Department, Aristotle University of Thessaloniki, 54124 Greece
\and
Academy of Athens, Research center for Astronomy and Applied Mathematics, Soranou Efesiou 4, 11527, Athens, Greece.
}

\offprints{Jorge Melnick \email{jmelnick@eso.org}}
\date{}

\authorrunning{Melnick et al.}

\titlerunning{The \lsigma relation in green}


\label{firstpage}
\abstract{
The correlation between emission-line luminosity ($L$) and profile width ($\sigma$) for HII Galaxies provides a powerful method to measure the distances to galaxies over a wide range of redshifts. In this paper we use SDSS spectrophotometry to explore the systematics of the correlation using the [OIII]5007 lines instead of \halfa\ or \hbeta\ to measure luminosities and line widths. We also examine possible systematic effects involved in measuring the profile-widths and the luminosities through different apertures. We find that the green \lsigma relation defined using [OIII]5007 luminosities is significantly more sensitive than \hbeta\ to the effects of age and the physical conditions of the nebulae, which more than offsets the advantage of the higher strength of the [OIII]5007 lines. We then explore the possibility of mixing [OIII]5007 profile-widths with SDSS \hbeta\ luminosities using the Hubble constant $H_0$ to quantify the possible systematic effects. We find the mixed \lhbeta$-\sigma_{[OIII]}$ relation to be at least as powerful as the canonical \lsigma relation as a distance estimator, and we show that the evolutionary corrections do not change the slope and the scatter of the correlation, and therefore, do not bias the \lsigma distance indicator at high redshifts. Locally, however, the luminosities of the Giant HII Regions that provide the zero-point calibrators are sensitive to evolutionary corrections and may bias the Hubble constant if their mean ages, as measured by the equivalent widths of \hbeta, are significantly different from the mean age of the HII Galaxies. Using a small sample of 16 ad-hoc zero point calibrators we obtain a value of $H_0=66.4^{+5.0}_{-4.5}~\kmsmpc$ for the Hubble constant,  which is fully consistent with the best modern determinations, and that is not biased by evolutionary corrections.
}
\maketitle

\keywords{HII galaxies; Starburst galaxies; Observational Cosmology}

\section{Introduction}

The relation between the integrated emission-line luminosity ($L$) and the velocity-width  ($\sigma$) of the emission-line profiles for HII Galaxies provides a powerful method for measuring the distances to these compact strong emission-line objects, and this out to redshifts of high cosmological relevance ($z>2$; \citealt{Chavez2016} and references therein).

In principle, the underlying astrophysics of the \lsigma relation should be surprisingly simple: HII Galaxies are powered by  young starburst clusters of ages of a few million years, so both the number of ionising photons and the turbulence of the gas are ultimately controlled by the total mass of the star-cluster and gas complex.

A first indication that this naive scenario may be incomplete is that the scatter in the \lsigma relation is substantially larger than the observational errors. Since the starburst clusters that ionise the nebulae evolve very rapidly in the first few million years, age is the natural culprit to explain the scatter. This effect has been well understood since the 1980's (eg. \citealt{Terlevich1981}) and was analysed in detail by \cite{Chavez2014}  who showed that luminosity evolution, as parametrised by the equivalent widths of \hbeta, {\em does not} explain the scatter in the relation. 

Instead, \cite{Chavez2014} showed that including the radius of the galaxy, $R$,  as a second parameter  reduces the scatter significantly as would be expected if HII Galaxies are Virialized systems. However, as pointed out by \cite{ter04}, HII Galaxies have complex star-formation histories, and many are also powered by more than one young starburst, so the physical meaning of the observed radii is difficult to interpret. 

In fact, a significant fraction of the objects in the sample compiled by \cite{Chavez2014} present multiple peaks in their emission-line profiles. Because of thermal broadening, these multiple profiles are often difficult to discern in the Balmer lines, but are clearly seen in the [OIII] lines, notably the green line at $\lambda$5007\AA, which in these objects have strengths comparable to \halfa\  and are significantly stronger that any other line in the spectrum. Moreover, for the highest redshift objects ($z>2.5$), [OIII]5007 is often the only line that can be measured with reasonable exposure times with state-of-the-art instrumentation on 8-10m class telescopes. 

For these reasons it is appealing to study the properties of the \lsigma relation as seen in the green light of the [OIII]5007 emission line, which is the aim of the present paper. In a parallel paper \citep{Arenas2016} we present the analysis of our most recent application of the ``canonical" \lsigma relation (i.e. using the Balmer lines) to measure the Hubble parameter ($H_0$) using a new sample of 37 Giant HII Regions in nearby galaxies as zero-point calibrators. In the present analysis we assume $\Omega_{\Lambda}=0.71$ and a flat Universe. $H_0$ is, of course, left as a free parameter, but when required as prior we adopt a value of $H_0=71~\kmsmpc$.

\section{The Data}
We have combined data sets from our previous works, to which we have added new measurements as described below.  

\subsection{Bordalo and Telles}

\cite{Bordalo2011} observed a sample of 120 local HII Galaxies (D$<$500Mpc) and we refer to the original paper for full details of the sample selection and the observational procedures. For the present purpose we recall that their velocity dispersions were determined with the FEROS  fibre-fed echelle spectrograph on the 1.52m and 2.2m telescopes on La Silla through a $2.7''$ entrance aperture. The fiber was positioned on the brightest knot of the star forming region (known as the kinematic core) that dominates the internal motions \citep{tel01,bor09}. The resolution of the spectra was $\mathcal{R}=48000$ corresponding to an instrumental velocity dispersion $\sigma_{inst} = 2.50\pm0.20$~\kms. 

The spectrophotometry was obtained with the Boller and Chivens spectrograph on the 1.52m telescope using a $2''$ slit \citep{keh04}. This information is relevant because we will be combining samples observed with different telescopes and different aperture sizes, and we must be aware of possible systematic effects. The present analysis is restricted to the best 70 objects from the sample of \cite{Bordalo2011}  { defined by V. Bordalo in his Ph.D. thesis using diagnostic diagrams to eliminate outliers}, which we will refer to as the BT11 sample.

As we will show below, the importance of including the BT11 sample is that it overlaps significantly in both luminosity and velocity dispersion with the Giant HII regions used as zero-point calibrators.


\subsection{Chavez et al.}
\cite{Chavez2014} studied a new sample of 128 HII galaxies: 122 selected from the SDSS on the basis of equivalent width of \hbeta\ (EW(H$\beta$) $>$ 50\AA) and redshift $z$ ($0.02 \leq z \leq 0.2$), the latter chosen to minimize the effects of local peculiar velocities in particular induced by the Local Supercluster and the Great Attractor, plus 6 galaxies in the southern hemisphere taken from the catalogue of \cite{Terlevich1991}. 

The emission-line profiles were obtained with two different spectrographs:  HDS on Subaru and UVES on the VLT, while the total emission-line fluxes and equivalent widths were measured from low-dispersion spectrophotometry obtained with the Mexican 2m telescopes at San Pedro Martir (Baja California) and Cananea (Sonora) through apertures larger than $8''$.  The HDS observations were taken through a $4''$ slit, which nominally corresponds to an instrumental velocity dispersion $\sigma_{inst}=12.3~$\kms, although the instrumental profile is flat-topped and not Gaussian. We will restrict our analysis to the subsample of 107 galaxies defined by  \cite{Chavez2014} as their best data set, which we will refer to as the Chavez14 sample.

\begin{figure}
\includegraphics[width=0.5\textwidth]{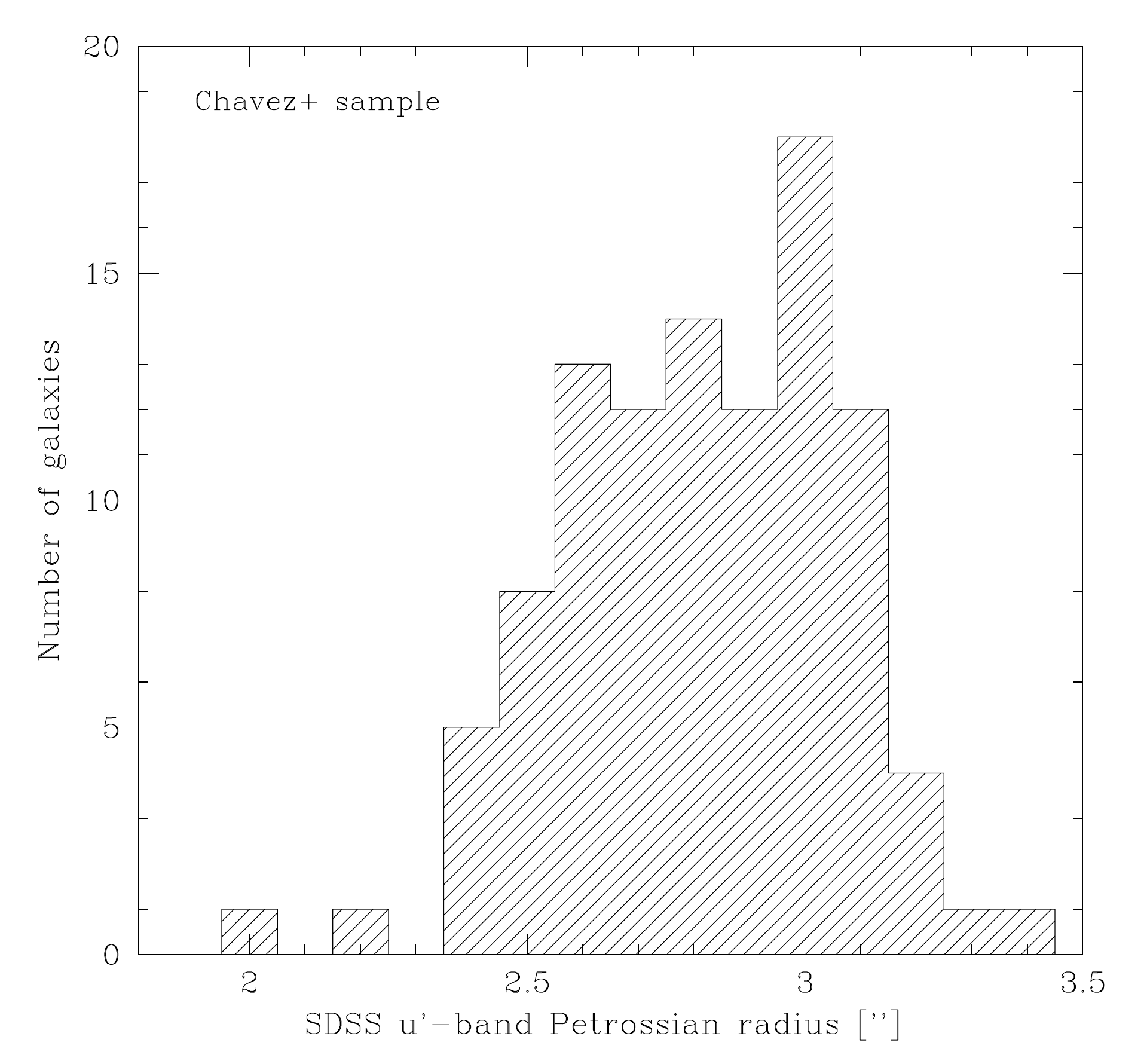} 
\vspace*{-0.0cm}
\caption{\small Distribution of SDSS Petrossian radii measured in the u'-band. }
\label{ruxu}
\end{figure}

As shown in Figure~\ref{ruxu}, the diameters of most objects in the Chavez14 sample are comparable to $4''$, so the instrumental resolution in the HDS spectra is determined not by the entrance slit but by the surface brightness profiles of the objects, which are very difficult to quantify using the available data.  \cite{Chavez2014} corrected their observations for instrumental broadening using a sort of ``equivalent" SDSS Petrossian diameter, { but since this procedure may introduce a distance-dependent error in the velocity dispersions, here we have chosen to use a different approach as discussed below.}

The UVES observations were obtained through a $2''$ slit corresponding to a nominal instrumental resolution of $\sigma_{inst}=4.65$~\kms, but again the instrumental profiles are also box-shaped. Although the seeing during the observations was substantially better than $2''$, the sizes of most objects are larger than the slit width, so no elaborate procedure was required for the instrumental corrections as was the case for the HDS data.  Notice, however, that both for UVES and HDS the flat-topped instrumental profiles imply that the line-profiles have non-Gaussian cores that can be clearly appreciated in the residuals of the Gaussian fits (\citealt{Chavez2014} and Appendix~A).

Following the earlier work of \cite{mel87}, \cite{Chavez2014} measured their photometric fluxes through wide apertures. However, as shown in Figure~\ref{ruxu}, the objects in the Chavez14 sample are quite compact, so the use of wide slits may introduce systematic aperture effects. Therefore, in order to combine the \cite{Chavez2014} and the \cite{Bordalo2011} samples, it seems safer to match as closely as possible the photometric and spectroscopic apertures, so for the Chavez2014 sample here we will use the spectrophotometry from SDSS taken through $3''$-diameter fibre apertures. 
 
{ Their "benchmark" sample (S3) of 107 galaxies, defined by \cite{Chavez2014} as objects with $log(\sigma)<1.8$~\kms}, contains 104 objects from the SDSS and 3 southern galaxies for which there is no SDSS data available. We downloaded the SDSS spectra  and re-measured all relevant line intensities, in particular [OII], [OIII], and the Balmer lines, using a dedicated pipeline that rebins the spectra to zero redshift and corrects the fluxes for foreground extinction using redshifts and extinctions provided by in the DR9 release of the SDSS.  We find that, even using the high-quality SDSS spectra available for these objects, the ratios of \hgama\ to \hbeta\ are much noisier than \halfa\ to \hbeta, so we have only used the latter to compute the internal extinction. We thus chose not to correct the \hbeta\ fluxes for underlying absorption, as was done by \cite{Chavez2014} using the so-called Q method.  

{ Our final catalog of Chavez14/SDSS objects is listed in Table~1 and the BT11 objects are listed in Table~2.} 
 
\subsection{Giant HII Regions}
 
It has been noted by several authors (e.g. \citealt{Hippelein1986,Bordalo2011}) that for HII Galaxies (HIIGs) and Giant HII Regions (GHIIRs), the Balmer lines are systematically broader than the [OIII] lines by about 2~\kms. Since the reasons for this difference are not well understood (see eg. \citealt{Hippelein1986}), for the purposes of calibrating the \lsigma relation using GHIIRs it seems safer to rely on the measured [OIII]5007 line-widths rather than blindly applying an empirical correction.\footnote{The correction is not so relevant for high-redshift objects that have $\sigma>40~$\kms, but is critical for the calibrating sample.}

Unfortunately, however, the high-resolution spectroscopic observations of both \cite{Chavez2012} and \cite{Arenas2016} used narrow-band order-separating filters to isolate \halfa, so we must rely on the scanning Fabry-Perot observations of \cite{Hippelein1986} that provide widths for both \halfa\  and [OIII]5007 for 21 GHIIRs for which spectrophotometry is available from the former authors. The Fabry-Perot data have the additional advantage that they are collected through wide apertures of diameters comparable to the spectrophotometric observations of \cite{Chavez2012} and \cite{Arenas2016}.  

The GHIIRs in NGC6822 have EW(H$\beta$)  lower than our threshold of 46\AA\ (see below), while those of NGC4256 lack equivalent width measurements in the literature, so we have excluded these 4 objects from the sample. We also eliminated one of the HII regions of M101, NGC5447, because it is a complex of several HII regions elongated along a spiral arm, and not a bonafide single Giant HII region. This leaves us with a sample of 16 zero point calibrators.

\subsection{Line profiles of HII Galaxies}

We already mentioned that most of the objects in our sample present some degree of non-gaussianity, probably due to multiple objects superimposed along the line of sight, to large expanding bubbles (which are actually clearly visible in HST images of local HII Galaxies), or both. 
{ In fact, 3D kinematical studies show that the brightest knots in HII Galaxies tend to have Gaussian profiles, while the more extended, low surface brightness regions have significantly broader profiles (see eg. \citealt{Moiseev2015} for a comprehensive discussion and references).
Thus, the integrated profiles are expected to have one or more dominant Gaussian cores and extended non-Gaussian wings.}

\cite{Bordalo2011} dealt with this problem by eliminating galaxies with clear multiple profiles on the basis of an analysis of the profile moments, while \cite{Chavez2014} resorted to fitting multiple Gaussians.  Examples of the range of typical [OIII]5007 line-profiles among the galaxies in our sample are presented in Appendix~\ref{apa}.

It seems risky, however, to eliminate objects that have multiple profiles because multiplicity will sometimes appear as elevated velocity dispersions and sometimes as elevated luminosities depending on the relative radial velocities of the starbursts that are superposed along the line of sight. Therefore, to avoid introducing subjective biases in our analysis of the \lsigma relation it seems safer to pay the price of including all the objects in the sample: a larger scatter.  

Still, we have eliminated two galaxies with ``flagrant" multiple profiles: J013258 and J142342 shown in Figure~\ref{profis}. The third `pathological' object in the figure, J212043, was not in the original Chavez14 sample. Thus, our final sample of Chavez14 with SDSS data - hereafter the Chavez14/SDSS sample - contains 102 galaxies.

In order to deal with non-Gaussian profiles we have chosen to use Gauss-Hermite functions that are commonly used for other problems in astronomy (eg. \citealt{rif10} and references therein).  Thus,  we have  fitted the original FEROS, HDS, and UVES  [OIII]5007 lines with Gauss-Hermite functions using the package PAN as implemented in IDL. Details of this can be found in \cite{wes07}. 

To minimise possible contamination by underlying old stellar populations, we restricted the sample to objects with equivalent widths, EW(H$\beta$)$>46$\AA,\  following the rationale explained by \cite{Chavez2014}.\footnote{The original selection criterion for the sample was EW(H$\beta$)$>50$\AA, but in our re-measured SDSS spectra the limit is actually 46\AA.} This reduces the sample to 32 objects from which we further eliminated two more: IIZw40, that is very extended, and UM559 that has an anomalous EW(H$\beta$) of 535\AA. This reduces our final sample of BT11 galaxies to 30 objects.  

\subsection{Errors}

\subsubsection{Line-profile widths}

Our treatment of the FEROS data is very similar to that of \cite{Bordalo2011} (having been done by the same person!), so we will retain the original errors that were evaluated from measurements of different lines of hydrogen and oxygen as a function of the S/N ratio as an internal test, and by measuring different spectra of the same object as an external test. 

The same is not possible for the Chavez14 data because the [OIII]5007 lines have significantly higher S/N than \hbeta\ so we need to propagate the errors including the corrections for instrumental ($\sigma_{ins}$) and thermal ($\sigma_{th}$) broadening. The thermal corrections are small for the oxygen lines, so can safely adopt a constant electron temperature of $T_e=12100{\rm K}\pm1440{\rm K}$ for all galaxies (that corresponds to the mean and standard deviation of the Chavez14/SDSS sample as measured by our pipeline). 

For the instrumental profiles we adopt: $\sigma_{\rm FEROS}=2.5\pm0.2$~\kms;  $\sigma_{\rm UVES}=4.65\pm2.32$~\kms; and $\sigma_{\rm HDS}=6.15\pm3.07$~\kms. This is equivalent to assuming that on average the objects in the Chavez14 sample have effective diameters of $r\simeq2''\pm 1''$. We do not have the same problem with the FEROS data because the spatial information is efficiently scrambled by the fibers. From the standard formula for the velocity dispersion, $\sigma=\sqrt{\sigma^2_{obs}-\sigma^2_{inst}-\sigma^2_{th}}$, the corresponding errors are calculated in the usual way by the square root of the sum of the variances in each variable.

Notice that our final velocity dispersion errors are significantly smaller than those of \cite{Chavez2014} for \hbeta. However, the [OIII]5007 lines are on average five times stronger than \hbeta\ and the errors become rather comparable if we multiply our values by $\sqrt{5}$.

\subsubsection{Luminosities}

For our Chavez14/SDSS sample the errors in luminosity are dominated by the errors in the fluxes and the extinction corrections. As mentioned before, we used only \halfa/\hbeta\ for the extinction corrections, whereas the errors in the fluxes were determined from the SDSS spectra using the statistical weights for each pixel given by the SDSS pipeline. For the BT11 data we use their indication that their fluxes, published by \cite{keh04}, have typical errors of 5\%,  which we propagate to the luminosities using the dispersion in the Balmer decrements from H$\alpha$ to H$\delta$ weighted by the corresponding line intensities to estimate the errors in the extinction.  As for the Chavez14 sample, we adopted the 30~Dor/SMC extinction law of \cite{Gordon2003}, thus,
  
\begin{equation}\label{uno}
 \delta logL(H\beta) = \sqrt{\Bigr(0.4343\frac{\delta F(H\beta)} {F(H\beta)} \Bigr)^2 +  (0.4\delta A_\beta)^2}
\end{equation}
where F(H$\beta$) is the observed flux, A$_\beta$ is the extinction at 4861\AA, and $\delta$ denotes the measurement errors in each variable. 

We recall that while the Chavez14 sample was selected to minimise distance errors due to local perturbations of the Hubble flow, so we can safely ignore distance errors, the \cite{Bordalo2011} sample contains many very local objects, so we are probably underestimating the errors in distance and hence luminosity.

\section{Results}

\subsection{The \lsigma relation in green}

\begin{figure}
\includegraphics[width=0.5\textwidth]{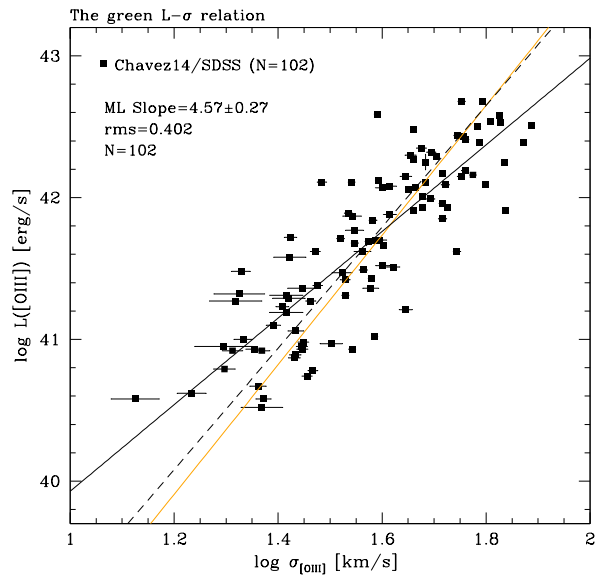} 
\vspace*{-1.5cm}
\caption{\small  The green $L-\sigma$~relation for our 102 HII galaxies with SDSS [OIII]5007 luminosities and Gauss-Hermite velocity dispersions. The black solid and dashed lines represent direct ($L$ vs. $\sigma$) and inverse ($\sigma$ vs. $L$) standard Least Squares (LSQ) fits to the data. The orange line shows a maximum-likelihood (ML) fit that takes the observational errors into account as described in Appendix~\ref{apb}. The relevant parameters of the three fits are listed in Table~\ref{rififi}. The 30 galaxies from the BT11 sample are not plotted because we lack [OIII]5007 luminosities.}
\label{green}
\end{figure}

We begin by examining the \lsigma relation for the [OIII]5007 line, which we will call the ``green'' \lsigma relation as opposed to the ``canonical" relation where we use \hbeta\ for both the luminosities and the velocity dispersions. The green relation is presented in Figure~\ref{green}.  The three lines correspond to the direct ($L$ vs. $\sigma$) and inverse ($\sigma$ vs. $L$) standard least-squares (LSQ) fits, both ignoring the observational errors, and to a maximum-likelihood (ML) fit that considers the observational errors in both variables.  For the latter we used an iterative algorithm originally proposed by Leon Lucy  (private communication) and described in detail in Appendix~\ref{apc}.  The relevant parameters for the three fits are listed in Table~\ref{rififi}.

Throughout this paper we will use the ML fit to define the slope of the \lsigma relation. Notice that, as expected since the errors in the independent variable ($\sigma$) are substantially larger than the errors in the dependent variable ($L$), the slope of the ML fit is very similar to the slope of the inverse LSQ fit.

The slope of the green \lsigma relation is  the same as that of the ``canonical" relation ($4.65\pm0.14$; \citealt{Chavez2014}), but steeper than the slope that we obtain using the SDSS \hbeta\ luminosities (see below).  The scatter of the green relation, however, is significantly larger than for \hbeta. \cite{Chavez2014} studied the sources of scatter in the \lsigma relation using multi-parameter fits and concluded that the radius of the HII Galaxies, measured on SDSS $u'$-band images, seems to be the dominant second parameter, as would be expected for virialized systems, and that age, metallicity, and excitation also contribute to the scatter.

The radius $R$ also appears to be the dominant second parameter for the green relation. Including log$R$ as a second parameter reduces the rms scatter to 0.22 (compared to 0.20 for \hbeta). \footnote{The multi-parameter fits use standard LSQ techniques that do not consider the observational errors, so the scatter cannot be directly compared to the scatter of the ML fits.}  The other two parameters that contribute significantly to the scatter of the green relation are age, as parametrised by the equivalent widths of \hbeta, EW(H$\beta$), and  { excitation, [OII]/[OIII]}. 

Thus, fitting a linear equation of the form,
\begin{equation}\label{rer}
\begin{split}
 log L([OIII]) = c_0 +c_{1}\times log \sigma_{[OIII]} + c_2\times logR \\
 +  c_{3}\times EW(H\beta) + c_{4}\times log ([OII]/[OIII])
\end{split}
\end{equation}
using standard LSQ techniques (i.e. ignoring the observational errors) yields 

\begin{equation}\label{coefs}
\begin{tabular}{l}
$c_{0} = 38.54\pm0.15$\\
$c_{1} = 1.64\pm0.11 $\\
$c_{2} = 0.80\pm0.06 $\\
$c_{3} = 0.0036\pm0.0004$ \\
$c_{4} = 0.14\pm0.04$
\end{tabular}
\end{equation}

Introducing the radius as a second parameter in the distance indicator is not practical because it not only weakens the distance sensitivity, but mostly because radii are difficult to measure even at low redshifts. Introducing the other two parameters is also not practical because both require measuring lines ([OII]3727 and \hbeta) that are substantially weaker than [OIII]5007 thus completely erasing the advantage of using [OIII] (using \hbeta\ luminosities has the additional advantage that the evolutionary signal ($c_{3}$) is significantly weaker than that of L([OIII]) ).
We conclude that because of its significantly larger scatter and sensitivity to additional parameters, the green \lsigma relation is not as useful as a distance indicator as the canonical relation.  In the next section we examine the possibility of combining [OIII]5007 line-profile widths with \hbeta\ luminosities.
   
\subsection{Back to \hbeta\ luminosities: the ``mixed" \lsigma relation}

\begin{figure}
\includegraphics[width=0.5\textwidth]{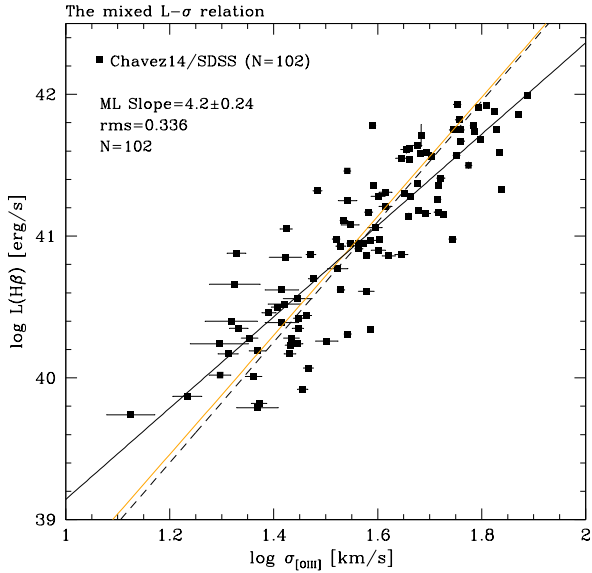}
\vspace*{-1.5cm}
\caption{\small  The mixed $L-\sigma$~relation for the Chavez/SDSS sample of 102 HII galaxies.  As in the previous figure, the lines represent the direct, inverse LSQ, and ML fits and the slope correspond to the value from the ML fit. We have not included  the 30 galaxies from our BT11 sample  to facilitate comparison with the green relation. The relevant parameters of the three fits are given in Table~\ref{rififi}.}
\label{beta}
\end{figure}

Figure~\ref{beta} shows SDSS \hbeta\ luminosities, \lbeta, plotted as a function of velocity dispersion from our Gauss-Hermite fits to the [OIII]5007 line profiles. The lines show the same three fits described in the previous section. We recall that the luminosities are obtained using our new measurements of the fluxes from the SDSS spectra, and therefore are not necessarily the same as those given by \cite{Chavez2014}.

The slope of the ``mixed" correlation for the Chavez14/SDSS sample is somewhat flatter than that of the green relation but the scatter is significantly lower. The evolutionary signal from Eq.\ref{rer} is significantly smaller than for [OIII]5007 luminosities: $c_3=0.0027\pm0.0004$, indicating that, as expected, the [OIII]5007 luminosities evolve faster than \lhbeta as the clusters age.  Curiously, the effect of excitation ([OII]/[OIII]), $c_4=0.25\pm0.04$, is stronger for \lhbeta than for L([OIII]).
 

\section{The Hubble Constant}

The purpose of this paper is not to measure the Hubble constant (in particular because we only have a restricted sample of Giant HII regions with [OIII] velocity dispersions), but rather to explore the systematics of the method. The determination of the Hubble constant using the \lsigma relation was discussed in \cite{Chavez2012} and will be revisited in a separate paper \citep{Arenas2016}. 

However, since we actually know the value of $H_0$ pretty accurately (\citealt{Riess2016, Arenas2016}), we can use the value of $H_0$ resulting from our relations as a parameter to quantify the systematics of the \lsigma relation.  We thus use the slope ($\alpha$) of the  \lsigma relation together with the data for our sample of 16 Giant HII regions with accurate Cepheid distances  to calibrate the zero point ($Z_p$) of the distance indicator as follows,

\begin{equation}\label{one}
Z_p=\frac{\sum_{i=1}^{16}{ {\it W_i}(logL_{\rm GHR,i} - \alpha\times log\sigma_{\rm GIIR,i}) }} { \rm \sum_{i=1}^{16}{{\it W_i} } }
\end{equation}
where $L_{\rm GHR,i}$ is the H$\beta$ luminosity of each Giant HII region and $\sigma_{\rm GHR,i}$ the corresponding velocity dispersion. The statistical weights $W_i$ are calculated as,

\begin{equation}\label{zero}
\begin{split}
W_i^{-1}= \Bigl(0.4343\frac{\delta L_{\rm GHR,i}}{L_{\rm GHR,i}}\Bigr)^2 + \Bigl(0.4343\alpha\frac{\delta\sigma_{\rm GHR,i}}{\sigma_{\rm GHR,i}}\Bigr)^2 \\
+ (\delta\alpha)^2(log\sigma_{\rm GHR,i}-<log\sigma_{\rm HIIG}>)^2
\end{split}
\end{equation} 
We stress that although the observational errors are not correlated, the covariance between $L$ and $\sigma$ must be considered in the propagation of errors as described in Appendix~\ref{apb}. Thus, from Eq.\ref{errp} here \mbox{$<log\sigma_{\rm HIIG}>$} is the average velocity dispersion of the sample of HII galaxies used to define the slope of the relation. 

The calibrated \lsigma relation is then: $ log L(H\beta) = \alpha log\sigma+Z_p$ and to calculate the Hubble constant we minimise the function, 

\begin{equation}\label{mins}
\chi^2(h) = \sum_{i=1}^{N}{W_i(\mu_i - \mu_{h,i})^2}
\end{equation}
where $\mu_i$ is the logarithmic distance to each HII galaxy calculated from the distance indicator and the reddening-corrected H$\beta$ fluxes F(H$\beta$) as,

\begin{equation}\label{mods}
2\mu_i=\alpha\times log\sigma_i + Z_p - logF_i(H\beta) - log(4\pi)
\end{equation}
and $\mu_{h,i}$ is the distance from the redshifts ($z_i$) using the standard cosmological equations for the luminosity-distance,

\begin{equation}\label{mod1}
\mu_{h,i} = log \Big[ D_{L,i}\equiv c(1+z_i)\int_0^{z_i} {\frac{dz'}{H(z')}}\Bigr]
\end{equation}
where the independent variable $h$ - the Hubble constant - is embedded in the equation for the luminosity distance, which we calculated using the analytical approximation of \cite{Wickra2010} assuming a flat universe and  $\Omega_{\Lambda}=0.71$.  This choice has a negligible effect in the distance moduli at the redshifts of our objects.

The best value of $h$ is then calculated minimising $\chi^2$ with statistical weights $W_i^{-1}=\delta\mu_i^2+\delta\mu_{h,i}^2$ calculated as,

\begin{equation}\label{weights}
\begin{split}
W_i^{-1}=(\delta Z_p)^2 + \Bigl(0.4343\frac{\delta F_i}{F_i}\Bigr)^2 + \Bigr(0.4343\alpha\frac{\delta\sigma_i}{\sigma_i}\Bigl)^2  \\
     +  (\delta \alpha)^2(log\sigma_i-<log\sigma>)^2
\end{split}
\end{equation}
 
\begin{figure}
\includegraphics[width=0.5\textwidth]{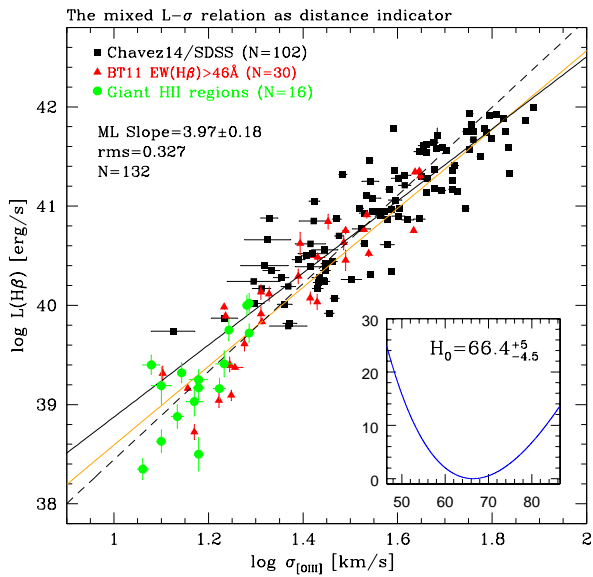}
\vspace*{-1.5cm}
\caption{\small  The mixed $L-\sigma$~relation for the combined Chavez14/SDSS (black squares) and BT11 sample (red triangles) of 132 HII galaxies. The lines show the LSQ and ML fits as in the previous figures and the relevant parameters are given in Table~\ref{rififi}. The green dots correspond to the 16 Giant HII Regions of the Hippelein/Arenas sample used to calibrate the zero point of the relation and thus establish the value of $H_0$. The inset shows the $\chi^2$ curve (actually $\chi^2-\chi_{min}^2$) and the value of $H_0$ determined from a cubic fit to the curve.}
\label{vinitu}
\end{figure}

Figure~\ref{vinitu} repeats Fig.~\ref{beta} but this time including the 30 objects from our BT11 sample, and the 16 Giant HII region from our Hippelein/Arenas sample for the determination of the zero point, and the corresponding value for the Hubble constant and $\chi^2$ curve.

We emphasise that we do not have a large enough sample of calibrator Giant HII Regions with [OIII] velocity dispersions, and therefore that our values for $H_0$ are only intended as a way to quantify the possible systematic effects involved in calibrating the \lsigma relation as a distance indicator.

\section{Systematics}
\subsection{Aperture matching}

From our initial studies in the 1980's our conventional wisdom has been to measure the velocity dispersions through relatively narrow slits, to preserve the spectral resolution, and the luminosities through wide apertures to include all the flux. The underlying assumption was that the turbulence is isotropic and the internal extinction modest, so that even through a narrow slit we still sample the full turbulent cascade. This is a pretty good assumption for single objects, but 50\% of our objects have complex profiles, and half of these show multiple peaks. Thus, even narrow slits encompass more than one starburst along the line of sight (see Figure~\ref{profis}), so using different entrance apertures for luminosities and velocity dispersions may introduce systematic effects. 

For this reason here we chose to use the SDSS spectrophotometry, that matches rather well the apertures used for the velocity dispersions  ($2'' -4''$), to measure luminosities. The fact that the galaxies in our high EW sub-sample of BT11 objects, observed through a $2''$ slit, are not systematically shifted relative to the Chavez14/SDSS sample despite being significantly more local, and that the scatter of the \lsigma relation in \cite{Chavez2014} -- using wide apertures -- is almost the same as that of our ``mixed" relation, indicates that the SDSS fluxes do not miss a significant fraction of the light.

{ Spatially resolved kinematic studies of HII Galaxies (e.g.. \citealt{Moiseev2012}) indicate that the integrated profile widths of these objects should in principle become broader as a function of aperture size. However, in practice the integrated profiles are dominated by the brightest cores of the galaxies, while the extended low surface-brightness regions appear as non-Gaussian wings that are effectively taken into account by our Gauss-Hermitte functions. } In fact, \cite{lag07}  showed using narrow-band imaging of some of the galaxies in the BT11 sample, that even in these local objects the luminosities increase only marginally for apertures larger than $2''-3''$, while the equivalent widths remain substantially unchanged.

\subsection{Age effects}

\begin{figure}
\includegraphics[width=0.5\textwidth]{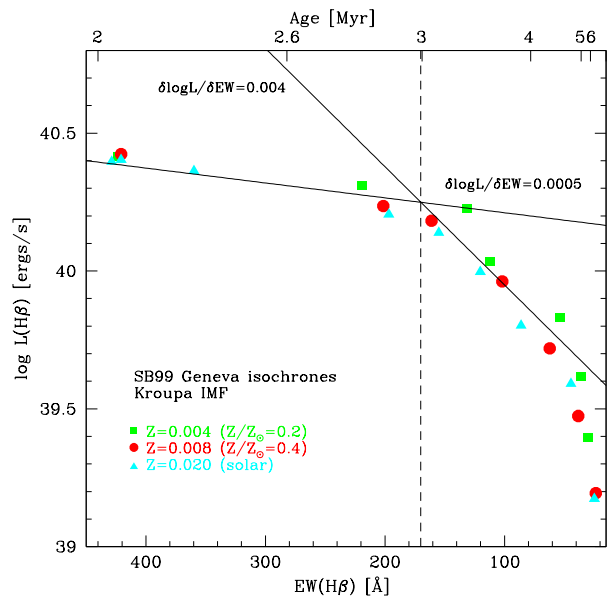} 
\vspace*{-0.0cm}
\caption{\small  The \hbeta\ luminosity evolution of single starburst clusters computed from Starburst99 models with parameters indicated in the legends. The lines show the fit for $\rm 45\AA<EW(H\beta)\le175\AA$ and $\rm EW(H\beta)>175\AA$. }
\label{sb99}
\end{figure}

The ionising fluxes of starburst clusters decrease steeply after about 3Myr, and thus their emission-line luminosities. The objects in our sample, on the other hand, span a range of ages, as estimated from the equivalent widths of the Balmer lines \citep{Leitherer1999} from 2.5 to 5 Myr, so in principle we must correct the \hbeta\ luminosities for evolution. Figure~\ref{sb99} presents the evolution of  \lbeta as a function of equivalent width EW(H$\beta$) computed for single starbursts using {\sc STARBURST99}  \citep{Leitherer1999} with Geneva isochrones without rotation and a standard Kroupa IMF, for a range of abundances spanning the values observed for our HIIGs. 

For all abundances the run of luminosity with equivalent width can be approximated reasonably well by two straight lines of slopes,

 \[
   \frac{dlogL}{dEW}=
   \begin{cases}
   	\rm
   	0.004 &   \text{if $ 45\AA<EW\le175$\AA} \\
	0.0005 & \text{if $EW> 175$\AA}
   \end{cases}
\]
over the range of equivalent widths spanned by our objects. 

Luminosity evolution, therefore, should leave a strong imprint in the scatter of the \lsigma relation. However, the turbulence of the gas may also evolve as the ionising clusters fade, but we lack a full understanding of the origin of the supersonic gas motions observed in GHIIRs and HIIGs.  The gravitational potential of the ionising cluster contributes a large fraction of the energy, but detailed studies of nearby objects, such as 30Doradus in the LMC and NGC604 in M33, show that gravity is only part of the story, and that hydrodynamic effects - winds, bubbles, filaments - also play an important role (e.g. \citealt{Medina1997,Melnick1999}).

Unfortunately, however, EW(H$\beta$) is not a precise age indicator for HII Galaxies \citep{ter04}. IR imaging by \cite{lag11} shows that HIIGs harbour many old and intermediate-age clusters, and in fact the majority of the galaxies in our sample show clear absorption components in \hdelta, which is the telltale of a significant intermediate-age stellar population. This is confirmed by population synthesis models of the broad-band spectral energy distributions (SED) of 238 SDSS HII Galaxies spanning more than two decades in wavelength (from the FUV at $0.15\mu m$ to the mid-IR at $22\mu m$;  \citealt{TellesMelnick2017}).

On the other hand, at least in principle GHIIRs should not be affected by this problem, and in fact correcting their observed luminosities for evolution using the models shown in Figure~\ref{sb99} leads to a significant reduction in the scatter of the \lsigma plane. Therefore, unless we correct both HIIGs and GHIIRs to the same fiducial age, we will introduce a systematic bias in the calibration of the zero point and thus in the value of the Hubble constant. 

The mean equivalent width of our sample of 16 calibrators is $<EW>=133$\AA, compared to $<EW>=95$\AA\ for our 132 HII galaxies. 
Preliminary results from our population synthesis models  \citep{TellesMelnick2017} indicate that, on average, the observed EW(H$\beta$) must be corrected by between 30\AA\ and 50\AA\ (depending on the details of the modelling) to account for the contribution of the underlying intermediate-age stellar populations to the continuum at \hbeta.  Thus, our best estimate for the average equivalent width of the starburst component(s) of our sample is  between 125\AA\ and 145\AA.

\begin{figure}
\includegraphics[width=0.5\textwidth]{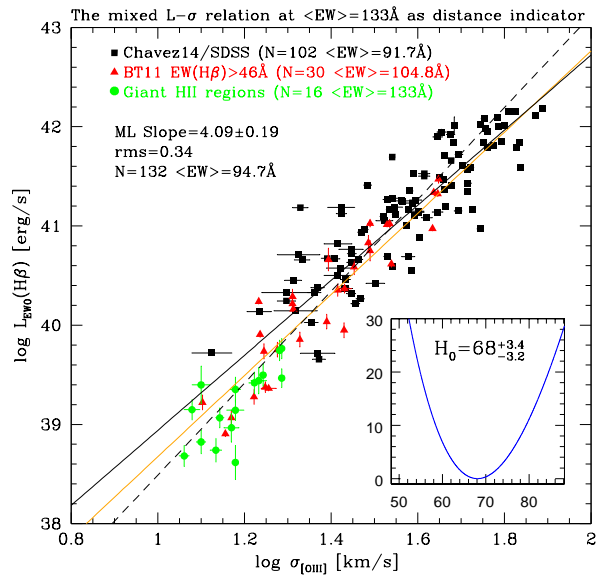} 
\vspace*{-1.5cm}
\caption{\small  The ``mixed'' L-$\sigma$ relation corrected for evolution at a fiducial ``age" of $<EW>=133$\AA\  (corresponding to about 3.5Myr for the Geneva isochrones). The remaining details of this figure are similar to those of Figure~4, except that here the mean value of EW(H$\beta$) is shown for each sub-sample and for the complete sample of HII Galaxies. The relevant parameters of the fits are listed in Table~\ref{rififi}. }
\label{3M}
\end{figure}

Figure~\ref{3M} shows the luminosities corrected to a fiducial ``age'' of $<EW>=133$\AA\ (corresponding to $\sim3.5$Myr for the Geneva models). The figure shows that the slope; the zero-point (ie. $H_0$); and the scatter of the \lsigma relation remain practically unchanged by the evolutionary corrections (see Table~\ref{rififi} for a comparison of the different slopes). However, if we correct the luminosities to the lower value of the mean corrected equivalent width of our HII Galaxies ($<EW>=125$\AA), we obtain $H_0=70.5^{+3.5}_{-3.3}~\kmsmpc$, while for $<EW>=145$\AA\ we find $H_0=64.3^{+3.2}_{-3.0}~\kmsmpc$, on average not significantly different from our uncorrected value of $H_0=66.4^{+5.0}_{-4.5}~\kmsmpc$ shown in Figure~\ref{vinitu}. The smaller errors of the corrected values are due to the reduced scatter of the zero-point calibrators (GHIIRs) resulting from the age-corrected luminosities, although these corrections introduce an additional uncertainty due to the correction of EW(H$\beta$) due to the contribution of the intermediate-age populations in the HII Galaxies.

Clearly, therefore, unless we are able to estimate accurately the ages of the starburst components of HII Galaxies, we will not be able to refine our estimation of $H_0$ to better than $\rm 4-5\ kms^{-1}Mpc^{-1}$.

\section {Conclusions}

We have studied the systematics of the \lsigma relation for HII Galaxies, and investigated the possibility of using the green lines of [OIII] at 5007\AA\ to define both the velocity dispersion ($\sigma$) and the luminosity ($L$) of the correlation. Our results can be summarised as follows:
\begin{itemize}

  \item The scatter of the \lsigma relation using only [OIII]5007 is substantially larger than for \lbeta\ mostly because the [OIII] luminosities of the starburst that power HII galaxies evolve faster than the \hbeta\ luminosities as the ionising stars age.  Thus, while the advantages of using [OIII]5007 for measuring $\sigma$ are strong, the same is not the case for the luminosities. The green \lsigma relation is not as good as the canonical \lsigma relation as a distance indicator; \\

  \item Combining \hbeta\ luminosities with [OIII] velocity dispersions provides the best compromise; the scatter of this ``mixed'' \lsigma relation is somewhat smaller than that of the "canonical" relation using \hbeta\ for both parameters, with the additional advantage that in HII Galaxies [OIII]5007 is significantly stronger than \hbeta\ and much less affected by thermal broadening;\\
  
 \item  Our "mixed" \lsigma relation established using Gauss-Hermite fits to the emission-line profiles, and SDSS spectro-photometry is at least as good, if not better, than the canonical relation as a distance indicator that uses wide-aperture photometry and a completely different technique to fit the line-profiles.  This demonstrates that the emission-line luminosities of HII Galaxies are dominated by the compact cores of these objects, and that the influence of the diffuse extended components is negligible;\\
 
 \item The scatter of the "mixed" \lsigma relation is substantially larger than the observational errors. This is partly due to the presence of one or more additional parameters in the relation. We confirmed the results of \cite{Chavez2014} that the radius, as measured on SDSS $u'$-band images, appears to be a second parameter of the correlation. We also found that starburst age, as measured by the equivalent-width of \hbeta\, EW(H$\beta$), has no influence on the scatter of the relation, contrary to what is expected from stellar-evolutionary models. However, EW(H$\beta$) is only an approximate age-indicator for HII Galaxies that contain massive intermediate-age populations, which contribute significantly to the continuum at \hbeta;\\
 
 \item Using an ad-hoc sample of 16 Giant HII Regions to determine the zero point, we used the calibrated "mixed" \lsigma relation for a sample of 132 HII Galaxies to measure the Hubble constant. Our value $H_0=66.4^{+5.0}_{4.5}~\kmsmpc$ agrees within the errors with the most recent determinations in the literature. We showed that these errors can be reduced by correcting the \hbeta\ luminosities of the zero-point calibrators (Giant HII Regions) for evolutionary dimming.  This, however, requires that similar corrections be applied to the HII Galaxies, which could only be done approximately. Our preliminary results suggest that our value for $H_0$  is probably not biased by age corrections, but this needs to be confirmed by detailed populations-synthesis modelling.
\end{itemize}
  
The main result from this investigation is that the \lsigma relation is robust to a number of potentially serious systematic biases, particularly aperture matching and non-Gaussian line profiles. While the scatter of the relation is larger than that of SNIa, for the purposes of measuring the Hubble constant, HII Galaxies are subject to different and independent systematics, while for measuring the expansion history of the Universe, HII Galaxies can be observed over a wider range of redshifts. The full power of the \lsigma relation to measure the Hubble constant will be presented in a forthcoming paper \citep{Arenas2016}.

\section*{Acknowledgements}
JM acknowledges support from a CNPq {\it Ci\^encia sem Fronteiras} grant at the Observatorio Nacional in Rio de Janeiro, and the hospitality of ON as a PVE visitor.  ET and JM acknowledge the hospitality of INAOE in Puebla during which parts of this research were conducted.

RC is grateful to the Mexican research council (CONACYT) for supporting this research under grant 263561.

Funding for the SDSS and SDSS-II has been provided by the Alfred P. Sloan Foundation, the Participating Institutions, the National Science Foundation, the U.S. Department of Energy, the National Aeronautics and Space Administration, the Japanese Monbukagakusho, the Max Planck Society, and the Higher Education Funding Council for England. The SDSS Web Site is http://www.sdss.org/.

The SDSS is managed by the Astrophysical Research Consortium for the Participating Institutions. The Participating Institutions 
are the American Museum of Natural History, Astrophysical Institute Potsdam, University of Basel, University of Cambridge, 
Case Western Reserve University, University of Chicago, Drexel University, Fermilab, the Institute for Advanced Study, the 
Japan Participation Group, Johns Hopkins University, the Joint Institute for Nuclear Astrophysics, the Kavli Institute for Particle
Astrophysics and Cosmology, the Korean Scientist Group, the Chinese Academy of Sciences (LAMOST), Los Alamos National
Laboratory, the Max-Planck-Institute for Astronomy (MPIA), the Max-Planck-Institute for Astrophysics (MPA), New Mexico State 
University, Ohio State University, University of Pittsburgh, University of Portsmouth, Princeton University, the United States 
Naval Observatory, and the University of Washington.

\bibliographystyle{aa}
\bibliography{mybib}
\label{lastpage}
\newpage

\onecolumn
\label{lobj}
\begin{longtab}
\tiny
\begin{longtable}{l l l l l}  
  \caption{\bf Objects from \cite{Chavez2014}} \\
  \hline\hline
Name &  redshift   & $ \sigma_{[OIII]5007}$ &  logL(H$\beta$) &  EW(H$\beta$) \\
  &  &  \ \ \ \ $km s^{-1}$& \ \ \ \ $erg s^{-1}$ &\ \ \ \ \  \AA \\
\hline
\endfirsthead                                                 
\caption{continues.}\\ 
\hline\hline                   
Name &  redshift   &   $ \sigma_{[OIII]5007}$ &   logL(H$\beta$)  &  EW(H$\beta$)\\
\hline
\endhead
        J000657 &   0.0737 & $     44.8 \pm    0.43$ & $    41.30\pm    0.01$ & $     86.5\pm     2.1$ \\ 
        J001647 &   0.0232 & $     21.3 \pm    0.89$ & $    40.88\pm    0.01$ & $     56.9\pm     0.7$ \\ 
        J002339 &   0.0531 & $     30.5 \pm    0.62$ & $    41.32\pm    0.01$ & $    110.1\pm     1.3$ \\ 
        J002425 &   0.0142 & $     34.9 \pm    0.54$ & $    40.31\pm    0.01$ & $     61.6\pm     0.8$ \\ 
        J003218 &   0.0180 & $     38.6 \pm    0.49$ & $    40.34\pm    0.01$ & $     80.0\pm     0.9$ \\ 
        J005147 &   0.0376 & $     25.6 \pm    0.74$ & $    40.50\pm    0.01$ & $     90.9\pm     1.1$ \\ 
        J005602 &   0.0582 & $     33.8 \pm    0.74$ & $    40.93\pm    0.01$ & $     48.7\pm     1.2$ \\ 
        J013344 &   0.0193 & $     17.1 \pm    1.10$ & $    39.87\pm    0.02$ & $     66.4\pm     1.8$ \\ 
        J014137 &   0.0181 & $     20.5 \pm    0.92$ & $    40.17\pm    0.01$ & $     62.7\pm     1.3$ \\ 
        J014707 &   0.0567 & $     53.2 \pm    0.36$ & $    41.15\pm    0.01$ & $    128.5\pm     1.6$ \\ 
        J021852 &   0.0128 & $     13.3 \pm    1.42$ & $    39.74\pm    0.01$ & $    138.3\pm     3.4$ \\ 
        J022037 &   0.1131 & $     50.6 \pm    0.42$ & $    41.56\pm    0.01$ & $    119.9\pm     2.0$ \\ 
        J024052 &   0.0824 & $     45.8 \pm    0.41$ & $    41.62\pm    0.01$ & $    274.3\pm     2.7$ \\ 
        J024453 &   0.0776 & $     38.2 \pm    0.67$ & $    41.17\pm    0.01$ & $     87.3\pm     1.7$ \\ 
        J025426 &   0.0148 & $     23.0 \pm    0.82$ & $    40.01\pm    0.01$ & $     53.2\pm     0.8$ \\ 
        J030321 &   0.1648 & $     67.3 \pm    0.40$ & $    41.75\pm    0.02$ & $    123.0\pm     3.4$ \\ 
        J031023 &   0.0515 & $     29.9 \pm    0.64$ & $    40.70\pm    0.01$ & $     67.6\pm     1.5$ \\ 
        J033526 &   0.0232 & $     23.4 \pm    0.84$ & $    40.19\pm    0.01$ & $     85.9\pm     1.7$ \\ 
        J040937 &   0.0748 & $     34.3 \pm    0.60$ & $    41.11\pm    0.01$ & $    118.6\pm     2.3$ \\ 
        J074806 &   0.0628 & $     34.9 \pm    1.44$ & $    41.25\pm    0.01$ & $    125.8\pm     2.5$ \\ 
        J074947 &   0.0742 & $     36.5 \pm    1.38$ & $    40.95\pm    0.02$ & $     57.4\pm     2.3$ \\ 
        J080000 &   0.0393 & $     26.5 \pm    1.90$ & $    40.85\pm    0.01$ & $     49.1\pm     0.9$ \\ 
        J080619 &   0.0698 & $     61.2 \pm    0.32$ & $    41.74\pm    0.01$ & $     68.5\pm     0.7$ \\ 
        J081334 &   0.0196 & $     29.0 \pm    0.65$ & $    40.44\pm    0.01$ & $     77.0\pm     0.9$ \\ 
        J081403 &   0.0199 & $     31.8 \pm    1.58$ & $    40.26\pm    0.01$ & $     92.6\pm     2.4$ \\ 
        J081420 &   0.0552 & $     36.6 \pm    0.66$ & $    40.91\pm    0.01$ & $     48.4\pm     1.1$ \\ 
        J081737 &   0.0236 & $     37.9 \pm    0.51$ & $    40.86\pm    0.01$ & $     56.8\pm     0.6$ \\ 
        J082520 &   0.0868 & $     45.7 \pm    0.51$ & $    41.14\pm    0.02$ & $     49.7\pm     2.1$ \\ 
        J082722 &   0.1086 & $     55.6 \pm    0.91$ & $    41.75\pm    0.02$ & $     65.4\pm     1.9$ \\ 
        J083946 &   0.1116 & $     56.5 \pm    0.89$ & $    41.57\pm    0.01$ & $     64.7\pm     1.3$ \\ 
        J084000 &   0.0723 & $     46.1 \pm    0.41$ & $    41.28\pm    0.01$ & $    153.2\pm     2.3$ \\ 
        J084029 &   0.0422 & $     39.1 \pm    0.48$ & $    41.36\pm    0.01$ & $    194.7\pm     2.3$ \\ 
        J084056 &   0.0504 & $     44.2 \pm    1.27$ & $    40.87\pm    0.01$ & $     65.3\pm     1.5$ \\ 
        J084219 &   0.0841 & $     44.2 \pm    1.14$ & $    41.55\pm    0.01$ & $     46.4\pm     0.7$ \\ 
        J084220 &   0.0295 & $     26.5 \pm    0.71$ & $    41.05\pm    0.01$ & $    114.6\pm     1.3$ \\ 
        J084414 &   0.0911 & $     56.6 \pm    0.91$ & $    41.93\pm    0.00$ & $     94.8\pm     0.8$ \\ 
        J084527 &   0.0311 & $     29.6 \pm    0.64$ & $    40.87\pm    0.01$ & $    116.3\pm     1.3$ \\ 
        J084634 &   0.0106 & $     27.1 \pm    0.70$ & $    40.23\pm    0.01$ & $     75.1\pm     0.7$ \\ 
        J085221 &   0.0760 & $     62.2 \pm    0.31$ & $    41.91\pm    0.00$ & $    148.9\pm     1.1$ \\ 
        J090418 &   0.0984 & $     60.8 \pm    0.83$ & $    41.78\pm    0.01$ & $     53.4\pm     0.9$ \\ 
        J090506 &   0.1256 & $     49.5 \pm    1.02$ & $    41.59\pm    0.01$ & $    100.2\pm     1.5$ \\ 
        J090531 &   0.0391 & $     37.9 \pm    1.33$ & $    40.61\pm    0.01$ & $    112.7\pm     1.3$ \\ 
        J091434 &   0.0273 & $     33.1 \pm    0.57$ & $    40.98\pm    0.00$ & $    104.0\pm     0.9$ \\ 
        J091640 &   0.0218 & $     27.2 \pm    0.96$ & $    40.28\pm    0.01$ & $    110.3\pm     1.6$ \\ 
        J091652 &   0.0570 & $     41.8 \pm    1.20$ & $    40.86\pm    0.01$ & $     77.0\pm     1.7$ \\ 
        J092540 &   0.0749 & $     52.0 \pm    0.97$ & $    41.17\pm    0.01$ & $     68.1\pm     1.8$ \\ 
        J092749 &   0.1070 & $     57.5 \pm    0.88$ & $    41.75\pm    0.01$ & $     83.0\pm     2.1$ \\ 
        J092918 &   0.0939 & $     39.9 \pm    1.26$ & $    41.28\pm    0.01$ & $    166.2\pm     3.9$ \\ 
        J093006 &   0.0136 & $     27.9 \pm    0.68$ & $    40.24\pm    0.01$ & $    113.3\pm     1.2$ \\ 
        J093424 &   0.0844 & $     52.6 \pm    0.96$ & $    41.41\pm    0.01$ & $     88.1\pm     1.5$ \\ 
        J093813 &   0.1021 & $     64.4 \pm    0.30$ & $    41.92\pm    0.01$ & $     74.1\pm     0.8$ \\ 
        J094000 &   0.0448 & $     39.9 \pm    1.26$ & $    40.90\pm    0.01$ & $     73.9\pm     1.4$ \\ 
        J094252 &   0.0149 & $     29.3 \pm    0.65$ & $    40.07\pm    0.01$ & $     84.0\pm     0.7$ \\ 
        J095000 &   0.0173 & $     27.0 \pm    0.70$ & $    40.17\pm    0.01$ & $     82.1\pm     1.2$ \\ 
        J095023 &   0.0977 & $     57.6 \pm    0.88$ & $    41.67\pm    0.01$ & $    101.6\pm     1.3$ \\ 
        J095226 &   0.1192 & $     57.1 \pm    0.89$ & $    41.82\pm    0.01$ & $     92.2\pm     1.7$ \\ 
        J095227 &   0.0150 & $     19.8 \pm    0.95$ & $    40.02\pm    0.01$ & $     76.8\pm     0.9$ \\ 
        J095545 &   0.0157 & $     28.1 \pm    0.68$ & $    40.42\pm    0.01$ & $     57.5\pm     0.9$ \\ 
        J100720 &   0.0314 & $     19.8 \pm    2.55$ & $    40.24\pm    0.01$ & $    110.6\pm     2.7$ \\ 
        J100746 &   0.0237 & $     33.8 \pm    0.57$ & $    40.62\pm    0.01$ & $    114.6\pm     1.3$ \\ 
        J101036 &   0.0395 & $     68.8 \pm    0.30$ & $    41.33\pm    0.01$ & $     68.1\pm     0.6$ \\ 
        J101042 &   0.0614 & $     45.8 \pm    0.41$ & $    41.54\pm    0.00$ & $     90.8\pm     0.7$ \\ 
        J101136 &   0.0547 & $     39.3 \pm    1.28$ & $    41.06\pm    0.01$ & $     83.0\pm     1.2$ \\ 
        J101430 &   0.1469 & $     66.8 \pm    0.76$ & $    41.88\pm    0.01$ & $     64.9\pm     1.7$ \\ 
        J102429 &   0.0333 & $     35.3 \pm    0.54$ & $    40.95\pm    0.01$ & $     91.7\pm     0.8$ \\ 
        J103328 &   0.0445 & $     62.9 \pm    0.33$ & $    41.68\pm    0.00$ & $     50.4\pm     0.4$ \\ 
        J103412 &   0.0687 & $     38.6 \pm    1.30$ & $    40.97\pm    0.02$ & $     67.9\pm     2.2$ \\ 
        J103726 &   0.0771 & $     41.2 \pm    1.22$ & $    41.21\pm    0.01$ & $     53.9\pm     1.4$ \\ 
        J104457 &   0.0132 & $     22.6 \pm    0.84$ & $    40.28\pm    0.01$ & $    266.7\pm     2.4$ \\ 
        J104554 &   0.0262 & $     40.1 \pm    0.48$ & $    40.98\pm    0.00$ & $    158.1\pm     1.4$ \\ 
        J104653 &   0.0107 & $     23.5 \pm    0.80$ & $    39.82\pm    0.01$ & $    173.9\pm     1.6$ \\ 
        J104723 &   0.0295 & $     47.5 \pm    0.40$ & $    41.37\pm    0.00$ & $     60.8\pm     0.5$ \\ 
        J104829 &   0.0927 & $     41.2 \pm    1.22$ & $    41.31\pm    0.02$ & $     85.1\pm     2.3$ \\ 
        J105032 &   0.0845 & $     38.9 \pm    0.49$ & $    41.78\pm    0.01$ & $    188.0\pm     2.1$ \\ 
        J105040 &   0.0523 & $     35.2 \pm    1.43$ & $    41.08\pm    0.01$ & $    105.3\pm     1.4$ \\ 
        J105210 &   0.1502 & $     45.2 \pm    1.12$ & $    41.61\pm    0.03$ & $     48.9\pm     2.3$ \\ 
        J105331 &   0.1238 & $     47.4 \pm    1.06$ & $    41.64\pm    0.02$ & $     62.3\pm     1.9$ \\ 
        J105741 &   0.0115 & $     28.6 \pm    0.66$ & $    39.92\pm    0.01$ & $     59.1\pm     0.7$ \\ 
        J110838 &   0.0238 & $     26.3 \pm    1.91$ & $    40.52\pm    0.01$ & $    119.2\pm     1.6$ \\ 
        J121329 &   0.0207 & $     27.9 \pm    1.80$ & $    40.56\pm    0.01$ & $     82.0\pm     1.0$ \\ 
        J130119 &   0.0692 & $     77.2 \pm    0.65$ & $    41.99\pm    0.00$ & $     85.5\pm     0.7$ \\ 
        J131235 &   0.0257 & $     26.0 \pm    1.93$ & $    40.62\pm    0.01$ & $     83.1\pm     1.1$ \\ 
        J132347 &   0.0225 & $     20.8 \pm    2.42$ & $    40.40\pm    0.01$ & $    246.9\pm     3.4$ \\ 
        J132549 &   0.0147 & $     26.1 \pm    1.93$ & $    40.39\pm    0.01$ & $    105.5\pm     1.1$ \\ 
        J134531 &   0.0304 & $     33.4 \pm    1.51$ & $    40.77\pm    0.01$ & $     62.2\pm     0.7$ \\ 
        J144805 &   0.0274 & $     49.3 \pm    1.02$ & $    41.16\pm    0.01$ & $    135.2\pm     1.3$ \\ 
        J162152 &   0.0344 & $     55.4 \pm    0.91$ & $    40.98\pm    0.01$ & $    135.6\pm     1.2$ \\ 
        J171236 &   0.0120 & $     23.4 \pm    2.15$ & $    39.79\pm    0.01$ & $    151.1\pm     1.6$ \\ 
        J210501 &   0.1428 & $     48.3 \pm    0.51$ & $    41.71\pm    0.08$ & $     56.4\pm     8.0$ \\ 
        J211527 &   0.0285 & $     21.1 \pm    2.38$ & $    40.66\pm    0.01$ & $    120.8\pm     1.6$ \\ 
        J211902 &   0.0896 & $     34.8 \pm    0.54$ & $    41.46\pm    0.01$ & $     74.4\pm     1.8$ \\ 
        J212332 &   0.0280 & $     21.5 \pm    0.89$ & $    40.35\pm    0.01$ & $     56.2\pm     1.4$ \\ 
        J214350 &   0.1098 & $     51.9 \pm    0.73$ & $    41.26\pm    0.02$ & $     55.3\pm     1.7$ \\ 
        J220802 &   0.1164 & $     68.3 \pm    0.31$ & $    41.59\pm    0.01$ & $     69.6\pm     1.4$ \\ 
        J221823 &   0.1084 & $     59.5 \pm    0.87$ & $    41.50\pm    0.02$ & $     54.9\pm     2.3$ \\ 
        J222510 &   0.0668 & $     52.0 \pm    0.38$ & $    41.36\pm    0.01$ & $    137.2\pm     2.0$ \\ 
        J224556 &   0.0805 & $     48.2 \pm    0.40$ & $    41.58\pm    0.01$ & $     67.5\pm     0.8$ \\ 
        J230117 &   0.0246 & $     24.6 \pm    0.77$ & $    40.46\pm    0.01$ & $     79.7\pm     1.2$ \\ 
        J230123 &   0.0304 & $     37.4 \pm    0.51$ & $    40.95\pm    0.01$ & $    120.1\pm     1.3$ \\ 
        J230703 &   0.1258 & $     74.4 \pm    0.29$ & $    41.86\pm    0.01$ & $     69.5\pm     1.1$ \\ 
        J231442 &   0.0342 & $     28.0 \pm    0.69$ & $    40.35\pm    0.02$ & $     62.4\pm     2.2$ \\ 
        J232936 &   0.0661 & $     47.6 \pm    0.40$ & $    41.18\pm    0.01$ & $     75.5\pm     1.5$ \\ 
 \hline
\end{longtable}
\end{longtab}

\newpage

\setcounter{table}{1}
\begin{longtab}
\tiny
\tabcolsep 2.0mm
\begin{longtable}{l l l l l}  
\caption{\bf Objects from \cite{Bordalo2011}}   \\
\hline\hline
Name &  redshift   &   $ \sigma_{[OIII]5007}$ &   logL(H$\beta$)  &  EW(H$\beta$)\tablefootmark{a}  \\
	  &		    &  \ \ \ \ $km s^{-1}$	& \ \ \ \ $erg s^{-1}$ &\ \ \ \ \  \AA \\
\hline
\endfirsthead                                    
\caption{continues.}\\         
\hline\hline                   
Name &  redshift   &   $ \sigma_{[OIII]5007}$  &   logL(H$\beta$) &  EW(H$\beta$) \\
\hline
\endhead
        UM382 &   0.0121 & $     18.0 \pm    0.72$ & $    39.37\pm    0.02$ & $    134.9\pm     1.5$ \\ 
        UM396 &   0.0208 & $     26.9 \pm    0.15$ & $    40.03\pm    0.07$ & $    153.1\pm     1.5$ \\ 
        UM417 &   0.0087 & $     14.8 \pm    0.17$ & $    38.72\pm    0.07$ & $     46.5\pm     1.5$ \\ 
      CTS1004 &   0.0473 & $     43.1 \pm    0.06$ & $    40.75\pm    0.02$ & $     77.3\pm     1.5$ \\ 
      CTS1005 &   0.0744 & $     43.4 \pm    0.10$ & $    41.34\pm    0.02$ & $    133.7\pm     1.5$ \\ 
      CTS1006 &   0.0207 & $     33.8 \pm    0.04$ & $    40.76\pm    0.02$ & $     69.8\pm     1.5$ \\ 
      CTS1008 &   0.0611 & $     44.3 \pm    0.38$ & $    41.35\pm    0.03$ & $    139.6\pm     1.5$ \\ 
  Tol0510-400 &   0.0413 & $     30.9 \pm    0.43$ & $    40.75\pm    0.03$ & $     64.0\pm     1.5$ \\ 
  Tol0633-415 &   0.0164 & $     30.6 \pm    0.06$ & $    40.63\pm    0.07$ & $     83.0\pm     1.5$ \\ 
 Cam0840+1201 &   0.0294 & $     34.3 \pm    0.04$ & $    40.91\pm    0.04$ & $    104.7\pm     1.5$ \\ 
  Tol1008-286 &   0.0138 & $     24.8 \pm    0.03$ & $    40.62\pm    0.11$ & $    122.7\pm     1.5$ \\ 
      CTS1011 &   0.0121 & $     20.5 \pm    0.09$ & $    40.13\pm    0.07$ & $     93.5\pm     1.5$ \\ 
      CTS1017 &   0.0354 & $     27.0 \pm    0.98$ & $    40.48\pm    0.04$ & $    161.1\pm     1.5$ \\ 
      CTS1018 &   0.0393 & $     30.9 \pm    0.19$ & $    40.45\pm    0.10$ & $     57.9\pm     1.5$ \\ 
      CTS1019 &   0.0665 & $     44.5 \pm    0.24$ & $    41.30\pm    0.04$ & $     90.4\pm     1.5$ \\ 
      CTS1020 &   0.0125 & $     34.6 \pm    0.05$ & $    40.52\pm    0.03$ & $    108.9\pm     1.5$ \\ 
      CTS1022 &   0.0137 & $     20.5 \pm    0.19$ & $    39.91\pm    0.07$ & $     56.9\pm     1.5$ \\ 
          F30 &   0.0034 & $     18.9 \pm    0.04$ & $    39.61\pm    0.07$ & $     97.3\pm     1.5$ \\ 
        MRK36 &   0.0021 & $     17.7 \pm    0.05$ & $    39.09\pm    0.05$ & $     61.7\pm     1.5$ \\ 
        UM439 &   0.0038 & $     17.6 \pm    0.04$ & $    39.40\pm    0.08$ & $     48.9\pm     1.5$ \\ 
        UM461 &   0.0035 & $     12.7 \pm    0.05$ & $    39.31\pm    0.07$ & $    155.2\pm     1.5$ \\ 
        UM463 &   0.0047 & $     16.7 \pm    0.06$ & $    39.04\pm    0.07$ & $     73.8\pm     1.5$ \\ 
      CTS1027 &   0.0067 & $     20.6 \pm    0.04$ & $    39.83\pm    0.02$ & $     50.0\pm     1.5$ \\ 
      MRK1318 &   0.0050 & $     17.1 \pm    0.05$ & $    39.98\pm    0.02$ & $     68.2\pm     1.5$ \\ 
  Tol1223-359 &   0.0093 & $     17.2 \pm    0.06$ & $    39.89\pm    0.03$ & $    128.8\pm     1.5$ \\ 
        UM570 &   0.0225 & $     21.3 \pm    0.12$ & $    40.11\pm    0.07$ & $    180.3\pm     1.5$ \\ 
       POX186 &   0.0042 & $     14.3 \pm    0.05$ & $    39.16\pm    0.03$ & $    274.2\pm     1.5$ \\ 
      CTS1035 &   0.0285 & $     26.0 \pm    0.15$ & $    40.07\pm    0.06$ & $     62.1\pm     1.5$ \\ 
 Cam1543+0907 &   0.0377 & $     28.4 \pm    0.05$ & $    40.84\pm    0.07$ & $    191.9\pm     1.5$ \\ 
  Tol2146-391 &   0.0295 & $     24.5 \pm    0.03$ & $    40.29\pm    0.07$ & $    245.5\pm     1.5$ \\ 
 \hline
\end{longtable}
\tablefoot{\tablefoottext{a}{Assumed errors}}
\end{longtab}

\twocolumn
\appendix
\section{Profiles}
\label{apa}

\begin{figure*}
\vspace*{-3cm}\includegraphics[width=1.08\textwidth,bb=100 0 1000 1000]{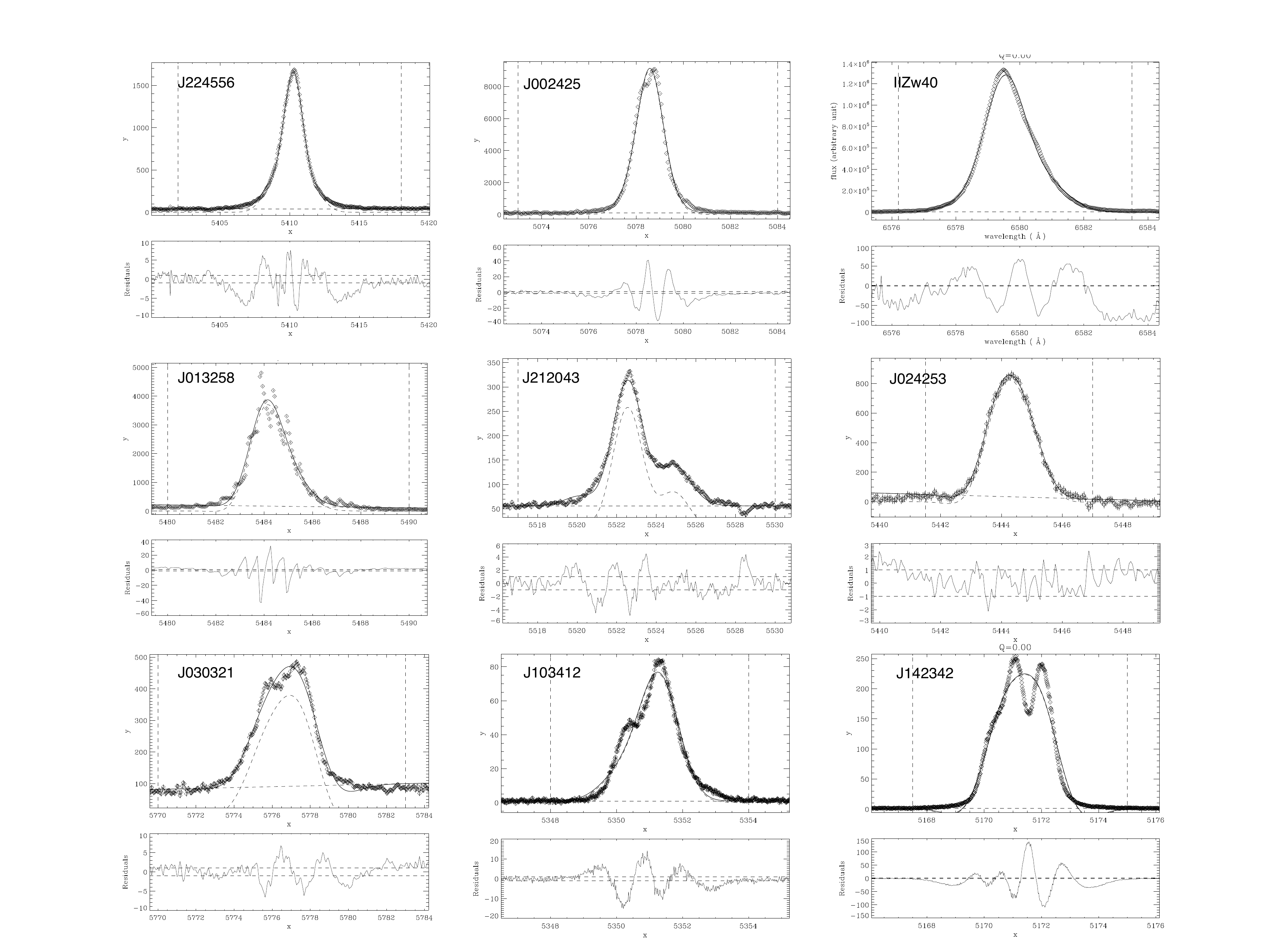} 
\caption{\small  Examples of the different emission-line profile shapes in our sample of HII Galaxies, together with the Gauss-Hermite fits. The lower panel shows the residuals from the fit.}
\label{profis}
\end{figure*}

Figure~\ref{profis} illustrates the zoo of emission-line shapes encountered in our sample of 132 HII Galaxies. The solid lines show the Gauss-Hermite profile-fits to the data, and the lower panel in each figure shows the residuals from the fit. The majority of the objects in our sample show some kind of profile asymmetry, be it extended wings or multiple components. Some cases, like J142342 in the lower-right corner of the figure, show two components of very similar strength, plus a third of slightly lower intensity. In this case we are integrating three bursts of comparable ages along the line of sight.

We have verified that, while in some cases the residuals in the \lsigma relation are correlated with complex profiles, this is not usually the case. Therefore, in our study we have made no attempt to select objects on the basis of the profile shapes. This clearly increases the scatter, but eliminates any subjective bias involved in the selection of acceptable profiles.

The figure also illustrates the usefulness of using [OIII]5007 instead of H$\beta$. The double structure of J002425 that is clearly discerned in [OIII]5007, is almost impossible to detect in H$\beta$ due to the larger thermal  broadening. It is also interesting to notice that the prototypical nearby HII Galaxy IIZw40, whose emission-line luminosity is dominated by a very compact core, shows a clear profile asymmetry even through the $2.7''$ FEROS entrance aperture. 

\section{Propagation of Errors}
\label{apb}
When we use an empirical relation, such as $L-\sigma$, to predict a value of the dependent variable we must include the covariance of the two variables when we propagate the errors \citep{Deming1964}. In general, the error  in the prediction of a linear correlation of the form
\begin{equation}\label{lin1}
y=a+bx
\end{equation}
at a given value of $x=x_m$,  $\delta Y_m$, when the parameters $a$ and $b$ are determined using least-squares techniques, can be expressed as \citep{Deming1964}

\begin{equation}\label{lin2}
(\delta Y_m)^2=\frac{(1-r^2)}{N-2}\sigma_y^2\Bigl(1+\frac{(x_m-<x>)^2}{\sigma_x^2}\Bigr)
\end{equation}
where $\sigma_x^2$ and $\sigma_y^2$ are the variances of the two variables, $<x>$ is the mean value of $x$, $N$ is the number of data points, and $r^2$ is the correlation coefficient defined as

\begin{equation}\label{lin3}
r^2=\frac{cov^2(x,y)}{var(x)*var(y)}
\end{equation}

For our particular application it is convenient to recall that for standard least-squares, the errors in the coefficients are given by \citep{Deming1964},

\begin{equation}
  \begin{tabular}{l}
  $(\delta a)^2 = \frac{(1-r^2)}{N-2}\sigma_y^2\Bigl(1+\frac{<x>}{\sigma^2_x}\Bigr) $\\
   $ (\delta b)^2 = \frac{(1-r^2)}{N-2}\frac{\sigma_y^2}{\sigma_x^2}$
    \end{tabular}
\end{equation}

so the error in the value predicted for a measurement $x_m$ can also be expressed as,
\begin{equation}\label{errp}
(\delta Y_m)^2 = (\delta a)^2 + (\delta b)^2(x_m^2-2x_m<x>)
\end{equation}
In order to minimise the effect of the covariance it is useful to replace the independent variable $x$ by $x'=x-<x>$ so that $<x'>=0$. This change of variables does not affect the error in the slope ($\delta b$), but does change the error in the zero point $\delta a$.  In cases like ours where the zero point $Z_p$ is determined from a set of independent calibrators (in our case giant HII regions), the covariance error is given by, 
\begin{equation}\label{errp2}
(\delta Y_m)^2 = (\delta Z_p)^2 + (\delta b)^2(x_m-<x>)^2
\end{equation}

When the data are subject to experimental errors in both variables ($\delta x_m,\delta y_m$), the error in the prediction becomes,  

\begin{equation}\label{errp3}
(\delta Y_m)^2 =(\delta y_m)^2 + (b\times\delta x_m)^2 + (\delta b)^2(x_m-<x>)^2+(\delta Z_p)^2 
\end{equation}
As we will discuss below, the standard least-squares solution is in general not unbiased when the independent variable is subject to error. In such cases Eq.\ref{errp3} is a very good approximation, but is not exact.

\section{The Maximum Likelihood Method}
\label{apc}

There is a substantial body of literature dealing with the question of how to fit straight lines to data affected by observational errors in both the dependent and the independent variables (see eg. \citealt{mac92,Tremaine2002}). For historical reasons, however, in this paper we have used the method originally suggested to us by Leon Lucy back in 1985, which is the one we used in the first papers dealing with the calibration of the $L-\sigma$ relation as a distance indicator \citep{mel87}.  In fact, here we actually used the same {\sc FORTRAN} code used in our early publications.

The central assumption of the method is that the error-free variables $(\xi_i,\eta_i)$ fit exactly a straight line of the form,
\begin{equation}\label{apaa1} 
\eta = m\xi + b
\end{equation}

The actual observations $(x_i,y_i)$ are subject to errors with standard deviations $(\sigma_{xi},\sigma_{yi})$, so the likelihood $\mathfrak{L}$ of the observational sample is,
\begin{equation}\label{apaa2} 
\mathfrak{L}=\prod_i \frac{1} {\sqrt{2\pi}\sigma_{xi} } e^ {-\frac{(x_i-\xi_i)^2} {2\sigma^2_{xi} }} \times \frac{1} {\sqrt{2\pi}\sigma_{yi} } e^ {-\frac{(y_i-\eta_i)^2} {2\sigma^2_{yi} } }
\end{equation}
or in logarithmic form,
\begin{equation}\label{apaa3} 
ln(\mathfrak{L}) = C - 0.5\sum{w_{xi}(x_i-\xi_i)^2} - 0.5\sum{w_{yi} (y_i-\eta_i)^2 }
\end{equation}
where $w_{xi}=1/\sigma^2_{xi}$ and $w_{yi}=1/\sigma^2_{yi}$  and C is a constant. Taking partial derivatives with respect to $m, b$, and $\xi$ we derive the equations,

\begin{equation}\label{apaa4}
\xi_i=\frac{ w_{xi}x_i+mw_{yi}(y_i-b)}{ w_{xi}+m^2w_{yi} }
\end{equation}

\begin{equation}\label{apaa5}
b=\frac {\sum{ w_{yi}(y_i-m\xi_i-b)} }  {\sum{w_{yi} }}
\end{equation}

\begin{equation}\label{apaa6}
m=\frac{ \sum{w_{yi}(y_i-b)\xi_i} }  {\sum{ w_{yi}\xi^2_i} } 
\end{equation}
which we solve by iteration. First guess the value of $m$ and $b$, usually by standard least-squares fits, and with these values estimate $\xi_i$ from equation~\ref{apaa4}. Then obtain new values for $m$ and $b$ using equations \ref{apaa5} and \ref{apaa6} and iterate the procedure until it converges.

To estimate the errors in the coefficients we make use of the basic formula for the variance of a quantity $\theta$,
\begin{equation}
var \hat{\theta}= \left[-\frac{d^2lnL}{d\theta^2}\right]^{-1}_{\hat{\theta}}
\end{equation}
where $\hat\theta$ is the Maximum-Likelihood estimate of the variable $\theta$.

\section{Parameters for the different fits}

The relevant parameters (slopes and rms) for the different linear fits employed in the text are presented in Table~\ref{rififi} to facilitate comparison between the different methods, and between the different parametrizations of the \lsigma relation as described in the text. 

The zero points for each fit are arbitrary and are not used in the determination of the Hubble constant, which is done using Giant HII Regions as zero-point calibrators.

\begin{table}
\caption{\bf Fit Parameters}
\tabcolsep 2.0mm
\tiny
\begin{tabular}{l c cc  cc  cc }  
\hline\hline
\lsigma			&&\multicolumn{2}{c}{ Max. Likelihood}	& \multicolumn{2}{c}{LSQ direct} 	& \multicolumn{2}{c}{LSQ inverse}	 \\
relation			& N & Slope & RMS 	       & Slope & RMS 		         & Slope & RMS 	\\ \hline
Green			&102 & $4.57\pm0.27$  & 0.402 & $3.05\pm0.19$ & 0.312 & $4.28\pm0.27$ & 0.397 \\
Mixed Chavez14	&102 & $4.20\pm0.24$  & 0.312 & $3.22\pm0.18$ & 0.287 & $4.25\pm0.24$ & 0.239 \\   		 
Mixed Chavez+BT11	&132 & $3.97\pm0.18$  & 0.327 & $3.63\pm0.15$ & 0.231 & $4.44\pm0.18$ & 0.337 \\
Mixed with Evolution	&132 & $4.09\pm0.19$  & 0.340 & $3.78\pm0.16$ & 0.320 & $4.63\pm0.19$ & 0.354 \\ \hline
\end{tabular}
\label{rififi}
\end{table}

\end{document}